# Can Centre Surround Model Explain the Enhancement of Visual Perception through Stochastic Resonance?

Ajanta Kundu
Microelectronics Division
Saha Institute of Nuclear Physics
1/AF Bidhannagar, Kolkata, India
ajanta.kundu@saha.ac.in

Sandip Sarkar [#]
Microelectronics Division
Saha Institute of Nuclear Physics
1/AF Bidhannagar, Kolkata, India
sandip.sarkar@saha.ac.in

*Abstract*—We demonstrate the ability of centre surround model for simulating the enhancement of contrast sensitivity through stochastic resonance observed in psychophysical experiments. We also show that this model could be used to simulate the contrast sensitivity function through stochastic resonance. The quality of the fit of measured contrast sensitivity function to the simulated data is very good.

## I. INTRODUCTION

Stochastic resonance (SR) is a phenomenon whereby small amount of additive noise can significantly enhance the performance of a non-linear signal processing system. The concept of stochastic resonance was first proposed by Nicolas, [1] and Benzi [2], to explain the near periodicity of the ice-ages, which coincides with periodic variation of earth's orbital eccentricity though this periodic force was too weak to cause such an abrupt change in earth's climate. The theory of SR proposed for the first time that an appropriate additive random noise may enhance the probability of detection of a non-linear sub-threshold signal. Any system consisting of (a) non-linearity (through barrier or threshold), (b) a sub-threshold signal, and (c) additive noise with a proper variance, are capable of exhibiting SR. There are many examples of SR in physical and biophysical systems such as, dithering system, Schmitt trigger, ring laser, Cray fish mechanoreceptor, cricket, human vision etc.

The idea of the association of noise with the nervous system is quite old. This led to the speculation of the positive role of noise in neural computation. It has been demonstrated in many experiments that the addition of external noise to a weak signal can enhance its detect ability by the peripheral nervous system of crayfish [3], cricket [4] and also human [5-8] by the process of SR. In all these experiments the neural recordings were analyzed, on the computer, for the presence of enhanced response through SR. All these were, therefore, indirect evidences of SR. It has also been demonstrated through psychophysical experiment [9] that human can make use of noise constructively for enhancing contrast sensitivity by the process of SR. It has been shown in this experiment that the brain can consistently and quantitatively interpret detail in a stationary image obscured with time varying noise and that both the noise intensity and its temporal characteristics strongly determine the perceived image quality.

It is well known that visual perception is a complex phenomenon involving higher level of cognition but it also includes lower level computation because for vision (visual computation) the very raw primal sketch is computed with the help of retina along with its associated circuitry. It is, therefore, expected that low-level computation (primal sketch) has a considerable role to play for observed enhanced visual perception by the process of SR.

What is then the role of low-level visual computation for the observed enhanced visual perception? Is it possible to build a low-level computational model that can exhibit enhanced contrast sensitivity by the process of SR? These are the issues we are trying to address in this work.

## II. BACKGROUND

The human retinal network consists mainly of three layers of cells, a two-dimensional array of primary photoreceptors, a layer of bipolar cells and a layer of ganglion cells. Information from rods and cones are being sent to the bipolar cells, either directly or through the network of horizontal cells. The bipolar cells, in their turn, send information to ganglion cells, either directly or through the network of amacrine cells. Information from ganglion cells go to cortex through visual pathway. Investigations [10-11] revealed that the image is extracted in successive layers through a "centre-surround" effect. This 'antagonistic' center-surround effect is modeled by difference of Gaussian or DOG [12-13] for which the resultant looks like a Mexican hat in two-dimension.

---

[#] Corresponding author

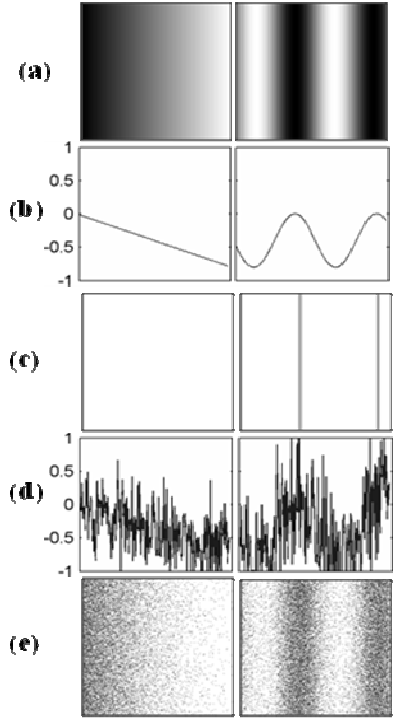

Fig. 1 Examples of zero-crossing maps of a ramp image and a sinusoidal grating (a) the original image (*I*), (b) profile of $I \otimes (-DOG - m\delta)$, (c) zero-crossing profile of the image in (b), (d) profile of $(I + noise) \otimes (-DOG - m\delta)$ and (e) zero-crossing map of the image in (d)

A DOG model in 2-D would be represented mathematically as:

$$DOG(x,y) = A_1 \frac{1}{\sqrt{2\pi}\sigma_1} e^{\frac{-(x^2+y^2)}{2\sigma_1^2}} - A_2 \frac{1}{\sqrt{2\pi}\sigma_2} e^{\frac{-(x^2+y^2)}{2\sigma_2^2}} \quad (1)$$

This model has been modified [14-15] to accommodate the concept of narrow channels [16] and the extended classical receptive field (ECRF)[17-20]. This is given by

$$-DOG(x,y) - m\delta(x,y), \quad (2)$$

where *m* is a constant factor and $\delta(x,y)$ is Dirac delta function in 2-D. There were claims of evidence in favor of zero crossing detector filters in the primary visual cortex [21-22]. This prompted us to use the above model for computing zero-crossing map of images for our investigation. An added advantage of the filter in (2) is that the zero-crossing map is capable of retaining shading information in the form of half-tone representation as shown in Fig. 1(e). It may be noted that even though a gray level image after zero crossing would be converted into a binary picture, having only two gray values for all the points, namely, either 0 (totally black) or 255 (totally white), the shading information of the original image is retained in the zero-crossing map.

### III. THE EXPERIMENT

To begin with we are investigating the usability of centre-surround model of retinal ganglion cells given by (2) for building a computational model capable of exhibiting SR. We would try to explain, with this model, some of the observations [9] related to the enhancement of contrast sensitivity with noise strength. We will also investigate whether this experiment can be extended to simulate "contrast sensitivity" curve similar to the one observed through the psychophysical experiment.

The image that we will be using for our investigation is a sinusoidal grating as shown in Fig. 2. The methodology of the simulation experiment is detailed below.

(a) We start with a synthetic image I as shown in Fig. 2 generated by $A\sin(2\pi fx) + 128$ and digitized on a 0-255 gray scale. Here, the amplitude *A*, denotes the contrast of the picture, *f* is the frequency and x is the spatial coordinate along which the pattern is changed.

(b) A random number *n* within 0-255, from a Gaussian distribution with zero mean and standard deviation $\sigma$ is added to original gray value *I* in every pixel so that every pixel value becomes $I + n$. Thus the noise in each pixel is incoherent with that of all other pixels but the standard deviation is the same for all.

(c) Next we choose a derivative filter function as in (2). We compute the derivative of the image by $(I+n) \otimes (-DOG - m\delta)$ where $\otimes$ denotes convolution. The resulting derivative image is bipolar and has pixels with negative as well as positive values.

(d) The zero-crossing map is then constructed from the resultant image in (c) by assigning a grayscale value 0 to each zero-crossing point and all other pixels in the image are assigned a value 255. This binary image resembles a half tone image where intensity variation of the original image maps to density variation of zero-crossing points. This is similar to the right image in Fig. 1(e).

(e) The steps (b)-(d) are repeated for various values of the contrast A. A typical example is shown in figure 3. The top most one is the original image and the rest of the images from top to bottom are zero-crossing images for increasing values of A. Even a visual inspection shows that the best reproduction of shading information is achieved

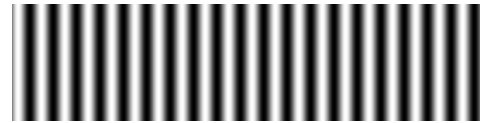

Fig. 2. Image of a typical sinusoidal grating used for the simulation study.

(third picture from the top) with moderate contrast A. This is a typical signature of SR.

(f) We now needs to have an estimate of the optimal contrast ($A_{opt}$) that will give best reproduction of the shading information of the original image. Taking into account the observations made by [23] we study the zero-crossing image in the Fourier domain and look for the minimum contrast ($A$) for which the second harmonic of the zero-crossing image just begins to appear. We designate this contrast as $A_{opt}$. This is the optimal contrast for the noise strength $\sigma$ for which the second harmonic of the zero-crossing image just begins to appear. For different values of the noise strength $\sigma$ different contrast (different $A_{opt}$) will be necessary for producing the above mentioned behavior of the second harmonic.

(g) We now repeat (a)-(f) for evaluating $A_{opt}$ for various values of the noise strength $\sigma$. The behavior of $A_{opt}$ with $\sigma$ is plotted in figure 4. It is evident from the figure that $A_{opt}$ is the minimum for an optimal amount of noise strength and increases for all other noise strengths. Alternatively contrast sensitivity ($1/A_{opt}$) increases most in the presence of an optimal amount of noise. Let us designate the minimum contrast as $A_{opt}^m$. This is a typical signature of SR observed in a psychophysical experiment by Simonotto [9].

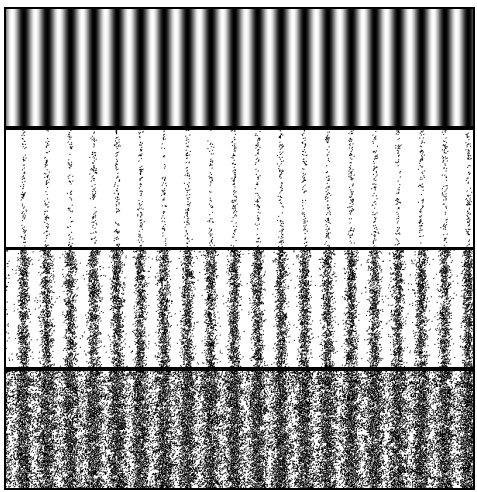

Fig. 3. Example of the effect of noise and contrast on the zero crossing. The topmost one is the original sine grating and the lower ones are the zero crossing images for increasing contrast for a given noise.

From the above study we find that (2) is capable of reproducing the dependence of the contrast sensitivity (1/contrast), through SR, on the amount of noise added to the original image. It also corroborates the finding that the contrast sensitivity attains a maximum for optimal noise strength. To validate our model given by (2) further, we try to reproduce the behaviour of the contrast sensitivity function (CSF) through SR. First we divide our visual frequency range by selecting an array of six isotropic two dimensional DOG filters whose center frequencies were arranged at octave intervals shown in Table 1 [24]. The ratio of center-surround space constants was 1:2. The spatial frequency bandwidth (full-width at half-height) was 1.9 octaves.

For each of the centre frequencies $f_c$ of these filters the above experiment, steps (a) – (g), were repeated with the grating image generated by $A\sin(2\pi f_c x) + 128$. We finally, therefore, get six $A_{opt}^m$ corresponding to six center frequencies. The plot of contrast sensitivity ($1/A_{opt}^m$) with the centre frequencies of the DOG filters is shown in figure 5. Each of the these points

Table1: Difference of Gaussian space constants

| Space constant (deg) | |
| --- | --- |
| Center | Surround |
| 0.0235 | 0.047 |
| 0.047 | 0.093 |
| 0.093 | 0.188 |
| 0.188 | 0.375 |
| 0.375 | 0.75 |
| 0.75 | 1.5 |

is the average of 20 iterations. For examining the goodness of fit with the measured contrast sensitivity function (CSF) [25] we have used

$$CSF(f) = K(0.0192 + 0.114f)\exp(-0.114f)^{1.1}, \quad (3)$$

to fit to the data using $K$ as the only adjustable parameter. The fitted curve is shown as solid line in Fig. 5. The quality of the fit of (3) to the simulated data is surprisingly good.

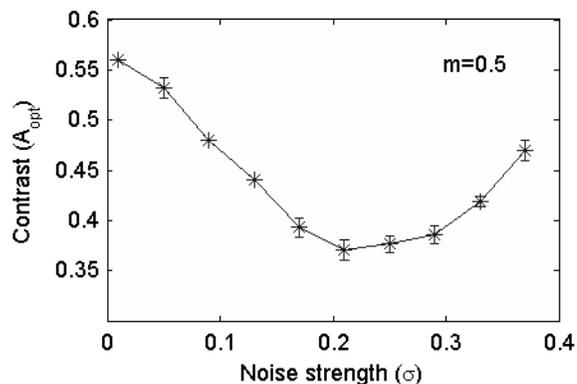

Fig. 4: Plot of optimum contrast with noise strength. The curve shows that the optimum contrast is minimum for an optimum amount of noise. Similar behaviour was also observed by Simonoto [9]

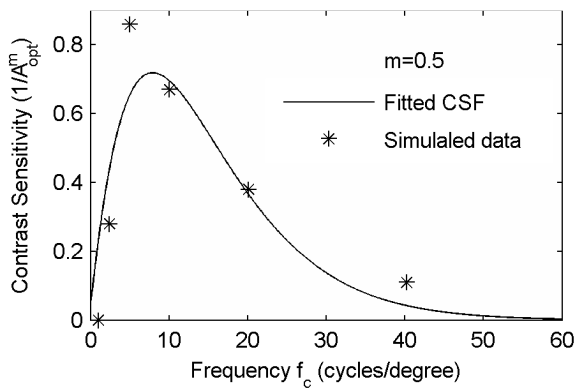

Fig. 5: Plot of contrast sensitivity ($1/A_{opt}^m$) with the centre frequency of DOG filters. The solid line is the fitted curve with the measured contrast sensitivity function (3)

## IV. RESULTS AND CONCLUSION

These experiments have demonstrated the utility of center surround model (2) for simulating some aspects of the visual system and its information processing in the presence of noise. The model could reproduce the nature of the enhancement of contrast sensitivity in the presence of optimal noise. The model could also simulate the contrast sensitivity function through SR. The quality of the fit of measured contrast sensitivity function to the simulated data is surprisingly good. The repeatability and stability of the model suggests that it may become a useful tool for understanding how our visual system interprets fine detail within noise contaminated images. This can also be used to study and build artificial system for enhancing or repairing contrast sensitivity in human.

ACKNOWLEDGMENT

We are grateful to Subhajit Karmakar for stimulating discussions and important suggestions.